\documentclass[9pt,twocolumn,twoside]{osajnl}
\journal{ol} % Choose journal (ao, aop, josaa, josab, ol, pr)

\setboolean{shortarticle}{true}
\title{All-passive multiple-place optical phase noise cancellation}

\author[1,*]{Liang Hu}
\author[1]{Ruimin Xue}
\author[1]{Xueyang Tian}
\author[1,2]{Guiling Wu}
\author[1,2]{Jianping Chen}

\affil[1]{State Key Laboratory of Advanced Optical Communication Systems and Networks, Department of Electronic Engineering, Shanghai Jiao Tong University, Shanghai 200240, China}
\affil[2]{Shanghai Key Laboratory of Navigation and Location-Based Services, Shanghai 200240, China}

\affil[*]{Corresponding author: liang.hu@sjtu.edu.cn}

\begin{abstract}
We report on the realization of delivering coherent optical frequency to multiple places based on passive phase noise cancellation over a bus topology fiber network.  This technique mitigates any active servo controller on the main fiber link and at arbitrary access places as opposed to the conventional technique, in which an active phase compensation circuit has to be adopted to stabilize the main fiber link.  Although the residual fiber phase noise power spectral density (PSD) in the proposed technique turns out to be a factor of 7 higher than that of in the conventional multiple-access technique when the access place is close to the end of the fiber link, it could largely suppress the phase noise introduced by the servo bumps,  improve the response speed and phase recovery time, and minimize hardware overhead in systems with many stations and connections without the need of the active servo circuits including phase discriminators and active compensators. The proposed technique could considerably simplify future efforts to make precise optical frequency signals available to many users, as required by some large-scale science experiments. 
\end{abstract}

\setboolean{displaycopyright}{true}

\begin{document}

%To date, efforts have mainly focused on long-distance connections between two locations connected by an optical fiber as first demonstrated in 1994 by Ma \textit{et al.} to correct phase perturbations between the two sites \cite{ma1994delivering}. 

\maketitle

Atomic optical clocks have rapidly grown in the last decade and are proving to be a powerful tool for investigation of fundamental and applied physics \cite{bloom2014optical, oelker2019demonstration, mcgrew2018atomic, huntemann2016single, brewer2019al+}. Precision clock networks are of particular interest in precision measurements and fundamental physics tests, such as general relativity, temporal variation of the fundamental constant \cite{parker2018measurement}, searching for dark matter, gravitational waves and physics beyond the Standard Model \cite{derevianko2014hunting, van2015search, arvanitaki2015searching, hu2017atom}, as well as providing innovative quantum technologies for other branches of science \cite{safronova2018search}. Most of the applications mentioned above require high-precision clock networks with multiple places over fiber links. {To achieve this aim, active compensation schemes as first demonstrated in 1994 by Ma  \textit{et al.} have been proposed to cancel the fiber-induced phase drift and implement highly stable optical frequency distribution \cite{ma1994delivering}, which generally utilizes the phase error from a round-trip probe signal to achieve the feedback control of a voltage-controlled oscillator (VCO) via a servo controller \cite{ma1994delivering}. One intriguing question is how to distribute a reference optical frequency to many users simultaneously in a cost-effective and robust way.}

\begin{figure*}[ht]
\centering
\includegraphics[width=0.92\linewidth]{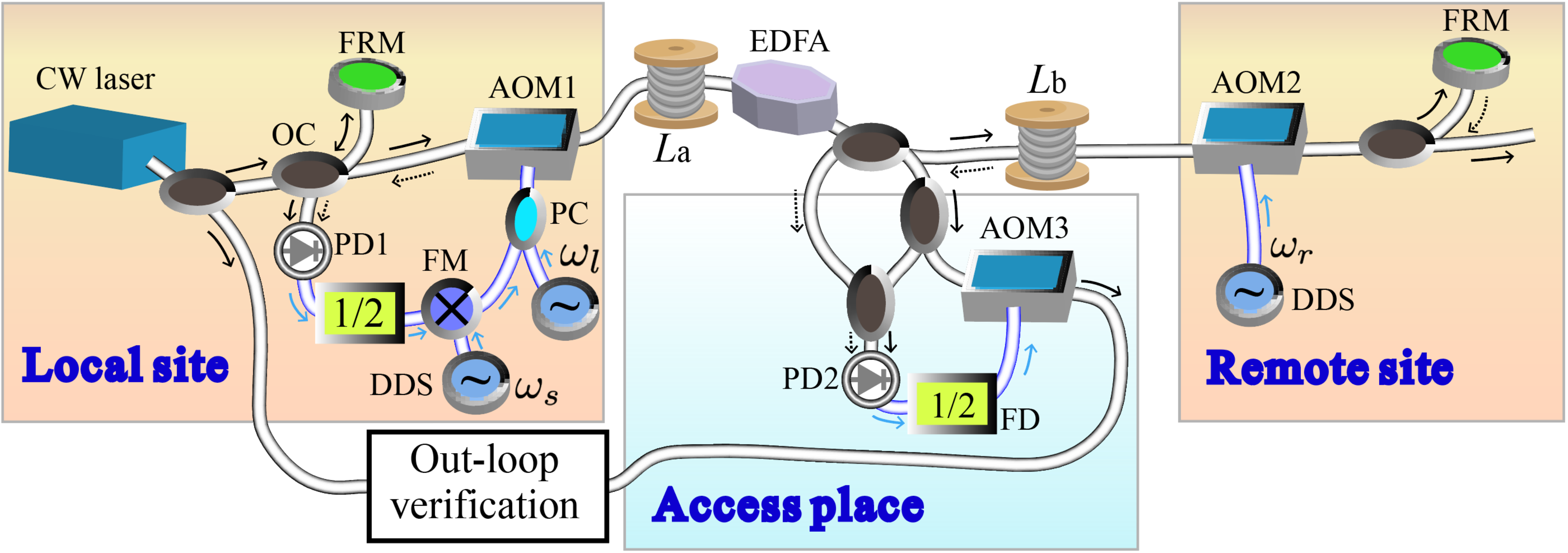}
\caption{Schematic diagram of our multi-access optical frequency dissemination with passive phase correction over a fiber link.  We evaluate the system performance by measuring the beatnote between the local light reference and the output of the access place. AOM: acousto-optic modulator, FRM: Faraday mirror, FM: frequency mixer, EDFA: erbium-doped fiber amplifier, OC: optical coupler, FD: frequency divider, PD: photo-detector, PC: power combiner, DDS: direct digital synthesizer. {The black solid and dashed arrows represent the directions of the forward and backward lights, respectively. The blue solid arrows are the direction of the RF signals.}}
\label{fig1}
\end{figure*}

In the past years, many works have demonstrated that coherence optical phase can be delivered to multiple users with the help of servo controllers.  Grosche \textit{et al.} have first proposed and demonstrated to extract the ultrastable signal for multiple users along the active phase stabilized main link \cite{grosche2014eavesdropping}, enabling to tap the fiber anywhere with the same precision level as that achieved at the main link output. Alternatively, a branching optical fiber network with phase noise correction at each output end has been proposed and experimentally demonstrated \cite{schediwy2013high, wu2016coherence}. All these existing demonstrators for the multiple-access applications have to adopt at least one active servo controller for the main fiber link \cite{bercy2014line, grosche2014eavesdropping, krehlik2013multipoint, grosche2010method} or for each branch fiber link \cite{schediwy2013high, wu2016coherence, hu2020multi}. Consequently, the overall performance of the above mentioned techniques is mainly dependent on whether the servo controller's  parameters set properly \cite{ang2005pid}. For example, there will be occasional interruptions where a phase lock temporarily loses lock, corresponding to unpredictable jumps in the optical signals. These interruptions will reduce the effective averaging time to bias the measurement \cite{ma1994delivering}. Moreover, the effect of the servo bumps on the main fiber link will pass to each access place and corrupt the spectral purity received by each place.

%\textcolor{red}{Although the active phase noise cancellation scheme can accomplish very high phase stability, the response speed and phase recovery time are restricted by the compensators’ parameters and optimization.} 
In our previous work, we have demonstrated optical frequency transfer with passive phase stabilization over a bus or a ring fiber network \cite{hu2020cancelling, hu2020multi}. The technique possesses the advantages of an unlimited compensation precision and a fast compensation speed and free from the effect of servo bumps on the spectral purity \cite{hu2020cancelling, hu2020multi}. However, the feasibility and adaptability of the passive phase noise cancellation technique for the multiple-access application needs to be theoretically and experimentally studied. Therefore, the investigation of the novel multiple-access optical frequency transfer scheme which implement alternative phase correction techniques other than the more commonly used conventional, active, phase correction technique is seeing increasing demand. Improving the robustness and sensitivity of multiple-access optical frequency transfer devices, as well as understanding and characterizing the limitations of the novel multiple-access optical frequency transfer scheme with passive phase correction is an important step towards the goal of heralding a new generation of viable precision optical frequency transfer devices. Theses devices could  be employed in the above mentioned applications \cite{bloom2014optical, oelker2019demonstration, mcgrew2018atomic, huntemann2016single, brewer2019al+, parker2018measurement, derevianko2014hunting, van2015search, arvanitaki2015searching, hu2017atom, safronova2018search}.

In this article, we extend the study of the optical frequency transfer technique based on our previous passive phase correction technique \cite{hu2020cancelling, hu2020multi}, demonstrating its application as a multiple-access optical frequency transfer technique.  We experimentally study the optical frequency transfer stabilities, phase noise power spectral densities (PSDs)  and accuracies for the two access places at the most symmetric 50/50 km ($L_a/L_b$) one and the relative asymmetric 70/30 km one, over a total fiber link of $L=100$ km. {In comparison with our previous work \cite{hu2020cancelling, hu2020multi, hu2020passive}, in which we have demonstrated multiple-place optical frequency transfer over the star and ring fiber networks with the passive phase noise cancellation technique, here we demonstrated that a high performance multiple-place optical frequency transfer over the widely adopted bus fiber topology with  all-passive phase noise cancellation at the access places and the main fiber link. The proposed technique could increase the adaptability to incorporate the optical frequency transfer technique into any existing communication networks with different topologies.}

Figure \ref{fig1} shows the schematic diagram of multiple-place optical frequency transfer with a simple extraction along the passively stabilized main link. The main optical link aims at regenerating a coherent optical phase at the output end of the fiber \cite{hu2020cancelling} and at any places along the fiber as the input end of the fiber. The principle of passively stabilizing the main fiber link can be found in \cite{hu2020cancelling, hu2020multi}. In brief, an ultrastable reference $\nu$ at the local site  is upshifted by frequency of $\omega_{l}$ with an acousto-optic modulator (AOM) denoted as AOM1 and then injects into the fiber. At the output, the light is downshifted by frequency of $\omega_r$ with the AOM2 (here $\omega_l>\omega_r$), and part of light is sent back with a Faraday mirror (FRM). The round-trip signal is mixed with the input ultrastable laser using an interferometer consisting of an optical coupler (OC) and another FRM. The beatnote radio frequency (RF) signal is twice the sum of the AOMs' frequencies and exhibits the round-trip phase noise, $2\phi_p$, with $\phi_{p}=\phi_{a}+\phi_b$, where $\phi_a$ and $\phi_b$ are the noise of the fiber spans of length of $L_a$ and $L_b$, respectively. Afterwards, we divide the beatnote with a factor of 2 and then mix the divided signal with an assistant frequency of $\omega_s$. The lower sideband of the mixed signal, $\omega_{s}-\omega_l+\omega_r$,  is used. Afterwards, the mixed signal together with $\omega_{l}$ is fed into the electrical port of the AOM1. Finally at the output end, the phase fluctuations of the optical frequency $\nu+\omega_{s}-\omega_l$ are automatically cancelled.

At each access place, the configuration is similar with the one presented in \cite{grosche2014eavesdropping}, at a distance $L_a$ from the input end and $L_b$ from the output end, a $2\times2$ optical coupler is adopted to extract both the forward and backward signals from the main fiber. The forward signal can be expressed as, 
\begin{equation}
\begin{split}
S_{F}(\omega)\propto&\cos((\nu+\omega_{l})t+\phi_a)\\
&+\cos((\nu+\omega_{s}-\omega_l+\omega_r)t-\phi_p+\phi_a).
\end{split}
\end{equation}

Similarly, the backward signal has a form of
\begin{equation}
\begin{split}
S_{B}(\omega)\propto&\cos((\nu-2\omega_r+\omega_{l})t+\phi_p+\phi_b)\\
&+\cos((\nu+\omega_{s}-\omega_l-\omega_r)t+\phi_b).
\end{split}
\end{equation}

 The beatnote frequencies between $\nu+\omega_l$ and $\nu-2\omega_{r}+\omega_l$ and between $\nu+\omega_{s}-\omega_l+\omega_r$ and $\nu+\omega_{s}-\omega_l-\omega_r$, result in the same frequency of $2\omega_r$ with the phase fluctuation $2\phi_b$. The signal is divided by a factor of 2, filtered, and drives an AOM (AOM3) in order to correct for the frequency and phase fluctuations of the forward signal.   The forward signal after passing through the AOM3, is thus downshifted to,
  \begin{equation}
\begin{split}
S_{F'}(\omega)\propto&\cos((\nu-\omega_r+\omega_{l})t+\phi_p)\\
&+\cos((\nu+\omega_s-\omega_l)t).
\end{split}
\end{equation}  
 
A similar phase noise cancelled signal can be obtained on the backward extracted signal with an opposite frequency shifter. Thus, the phase noise of the remote site and the access sites are automatically cancelled.  

{As demonstrated by Williams \textit{et al.},  the phase noise rejection capability is limited by the propagation delay \cite{williams2008high}. The residual phase noise PSD, $S_{\text{access}}(f)$, at each access place in terms of the single-pass free-running phase noise PSD, $S_{\text{fiber}}(f)$, and the single-pass fiber link $L$ propagation delay, $\tau_0$, by compensating the forward optical signal can be expressed  as \cite{williams2008high, hu2020passive},
\begin{equation}
S_{\text{access}}(f)\simeq{(2\pi f\tau_0)^2}\left[1+2\frac{L_a}{L}-\frac{2}{3}\left(\frac{L_a}{L}\right)^2\right]S_{\text{fiber}}(f).
\label{eq4}
\end{equation}}

\begin{figure}[htbp]
\centering
\includegraphics[width=0.95\linewidth]{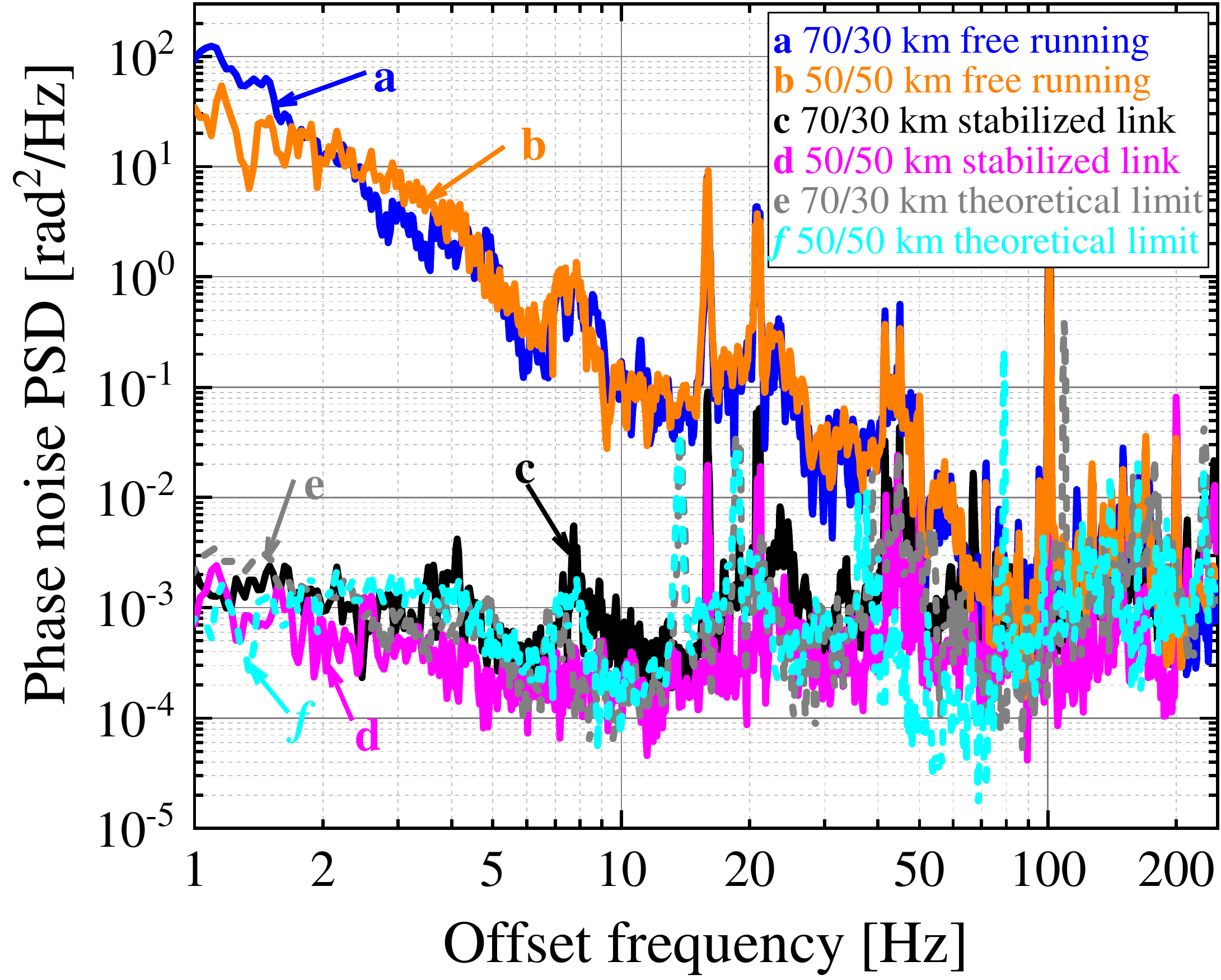}
\caption{Measured phase noise PSDs of \textbf{a} the 70/30 km free-running access place (blue curve), \textbf{b}  the 50/50 km free-running access place (orange curve), \textbf{c}  the compensated 70/30 km access place (black curve), \textbf{d}  the compensated 50/50 km access place (magenta curve), \textbf{e}  the 70/30 km theoretical delay-limited value (gray curve), and \textbf{f}  the 50/50 km theoretical delay-limited value (cyan  curve). Strong servo bumps can be effectively removed in the proposed  scheme.}
\label{fig2}
\end{figure}

Note that, in comparison with the conventional multiple-access technique, the phase noise rejection capability of both the conventional and proposed techniques is proportional to $\tau^2_{0}$ and the uncompensated single-pass fiber phase noise PSD, $S_{\text{fiber}}(\omega)$. Although, the residual fiber phase noise PSD in the proposed technique turns out to be a factor of 7 worse than that of in the conventional multiple-access scheme at the output of the fiber link ($L_a=L$) \cite{bercy2014line, hu2020multi}, it could largely suppress the phase noise introduced by the servo bumps (see Fig. \ref{fig2}), which could significantly increase as the increase of the fiber link \cite{raupach2015brillouin}, and  simplify the hardware overhead in systems with many stations and connections \cite{hu2020cancelling}.

We implement the proposed extraction scheme in Fig. \ref{fig1} on a 100 km fiber link.  The signal source that has been adopted in this experiment is a narrow-linewidth optical source (NKT X15) at a frequency of near 193 THz with a typical linewidth of 100 Hz. At an arbitrary access place,  a $2\times2$ optical coupler (coupling ratio 50\%: 50\%) is used to tap both forward and backward lights. To demonstrate the phase noise rejection capability at any places along the fiber link, we choose two representative places, which are at the most symmetric one, 50/50 km ($L_a/L_b$), and the relative asymmetric one, 70/30 km. To compensate the total loss of 23 dB of the 100 km fiber link system, we implement a home-made bidirectional Erbium-doped fiber amplifier (EDFA) at the middle of the fiber link.  The angular frequencies at the local and remote sites are $\omega_{l}=2\pi\times75$ MHz, $\omega_{r}=2\pi\times45$ MHz and $\omega_{s}=2\pi\times115$ MHz, respectively. The AOM1 is simultaneously supplied by two angular frequencies of $2\pi\times85$ MHz and $2\pi\times75$ MHz,  and the AOM3 is supplied  by the angular frequency of $2\pi\times45$ MHz. Consequently, the beatnotes between the input and extracted sites have a frequency of $2\pi\times40$ MHz. This beatnote is recorded simultaneously with a dead-time free counter with a gate time of 1 s and non-averaging $\Pi$-type operation. At the same time, we implement the phase noise measurement by feeding the heterodyne beatnote signal together with a stable frequency reference to a phase detector. The phase fluctuations $S_{\phi}(\omega)$ at the phase detector output are measured with a fast Fourier transform (FFT) analyzer \cite{hu2020passive, tian2020hybrid}.

\begin{figure}[htbp]
\centering
\includegraphics[width=0.95\linewidth]{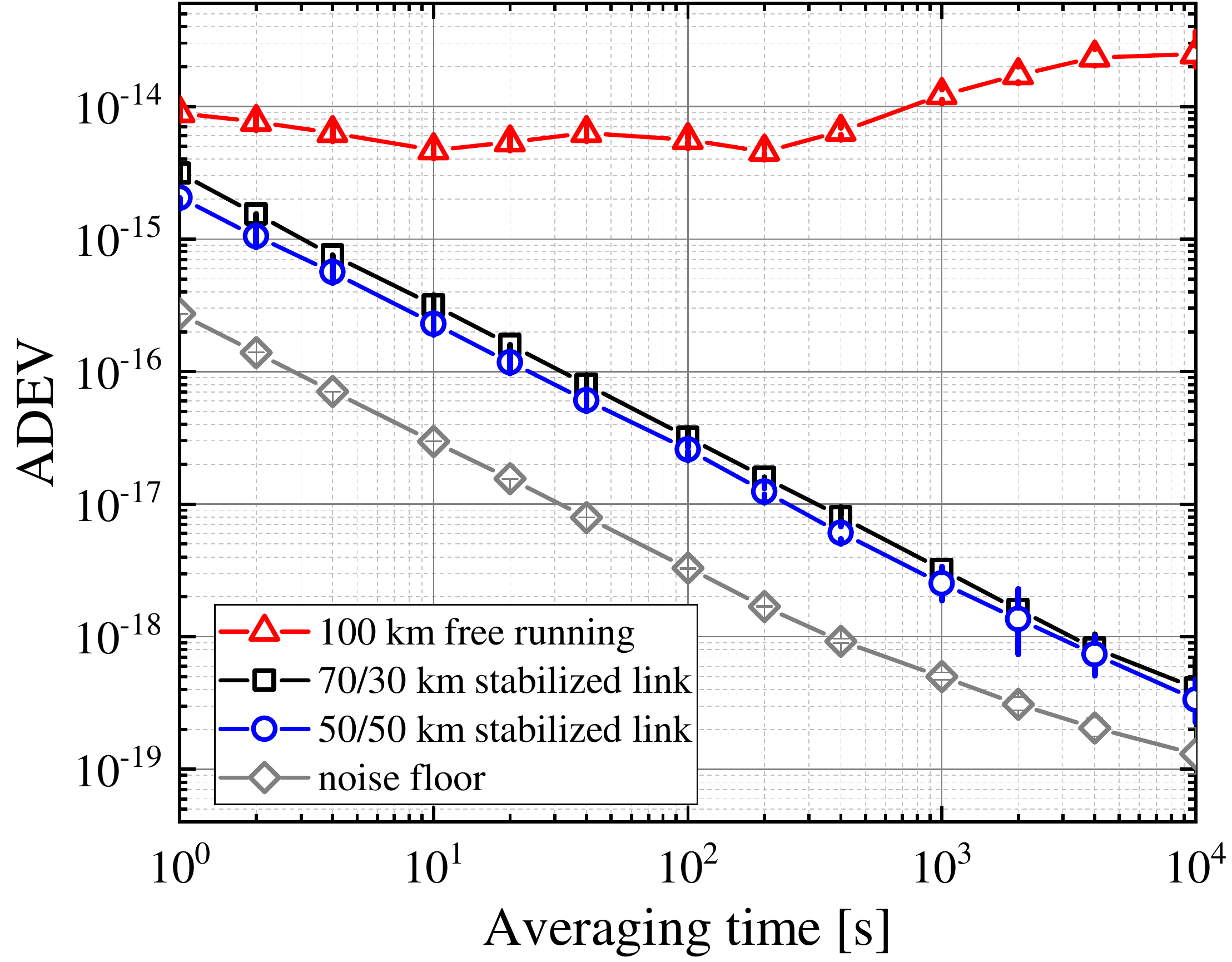}
\caption{Measured fractional frequency instability, calculated from $\Pi$-type data with the overlapping Allan deviation (ADEV), of the 100-km free-running
main link (red  triangles), the stabilized 70/30 km access place output (black squares),  the compensated 50/50 km access place output (blue cicles), and the noise floor, in which the fiber links are replaced by the short fibers, of the access place output (gray diamonds)}
\label{fig3}
\end{figure}

Figure \ref{fig2} shows the phase noise power spectral densities (PSDs) of the stabilized 70/30 km access place  (\textbf{c}, black curve) and the stabilized 50/50 km access place (\textbf{d}, magenta curve), respectively. The phase noise PSDs of the free-running 70/30 km access place (\textbf{a}, blue curve) and the 50/50 km access place (\textbf{b}, orange curve) are also shown.  The free-running curve is typical for optical fibre links, with a noise of approximately $100$ rad$^2/$Hz at 1 Hz for the 70/30 km access place, scaling down with a slope of about $f^{-2}$, and reaching around $10^{-3}$  rad$^2/$Hz after 100 Hz. Their noises of the free-running 70/30 km (\textbf{a}, blue curve) and 50/50 km access places (\textbf{b}, orange curve) slightly differ because the two measurements are performed at different times. At the same time, both passive phase stabilized cases  are also very similar with the phase noise PSDs about $10^{-3}$ rad$^2/$Hz between $1$ and $100$ Hz. We can clearly see that the noise correction is limited by the main fiber propagation delay approximately $\sim150$ Hz, which is compatible with the theoretical bandwidth of 189 Hz given by $1/(4\sqrt{\text{7}}\tau_0)$. This limit is the same for both access places because the bandwidth of the extraction is limited by the longer delay, namely the main fiber link delay \cite{bercy2014line}. More importantly, we can clearly see that compared with the conventional multiple-access optical frequency transfer technique \cite{grosche2014eavesdropping, schediwy2013high, wu2016coherence}, the strong servo bumps are effectively suppressed in our passive phase noise cancellation technique as observed in our previous work \cite{hu2020multi, hu2020cancelling, hu2020passive}.

%the noise rejection of around $10^6$ at 1 Hz is also compatible with the theoretical limit given by Eq. \ref{eq4}, which is $xxx\times10^{7}$ for a 100 km link. 

%with a correction overshoot of a few hundreds of Hz

% which is close to the theoretical calculation. As introduced above, the passive noise correction is very robust and that the set-up was operated during half day without any cycle slip due to the absence of any servo circuits to dynamically extract and  compensate the phase noise. The access place indeed exhibits both the contribution of the signal extracted from the main link. 

Figure \ref{fig3} displays the fractional frequency stability of the free-running and passively stabilized configurations, calculated from $\Pi$-type data with overlapping Allan deviation (ADEV).  The  stability of the 70/30 km (50/50 km) access place is $3.1\times10^{-15}$ ($2.0\times10^{-15}$) at 1 s averaging time, decreases as a slope of approximately $\tau^{-1}$ and reaches a floor of approximately $4.1\times10^{-19}$ ($3.3\times10^{-19}$) at $10^4$ s.  As calculated by Eq. \ref{eq4}, the ratio of the stability of the $50/50$ km and $30/70$ km places should be $\mathcal{R}=0.64$. In our experiment, we also obtain the ratio of $\mathcal{R}\simeq0.64$ ($2.0\times10^{-15}/3.1\times10^{-15}$). As a comparison, we also measured the noise floor for  the access output as displayed in Fig. \ref{fig3}. The long-term stability is mainly attributed to the thermal noise on uncompensated fibre paths, due to imperfect length adjustment and thermal stabilisation in the extraction optical set-up, or in the main fiber link \cite{hu2020passive, tian2020hybrid}. Compared with our previous work in which we characterized the system performance at the output of the main fiber link \cite{hu2020cancelling}, the ultrastable laser can be thus transferred through the secondary and main links without significant degradation.

\begin{figure}[htbp]
\centering
\includegraphics[width=1\linewidth]{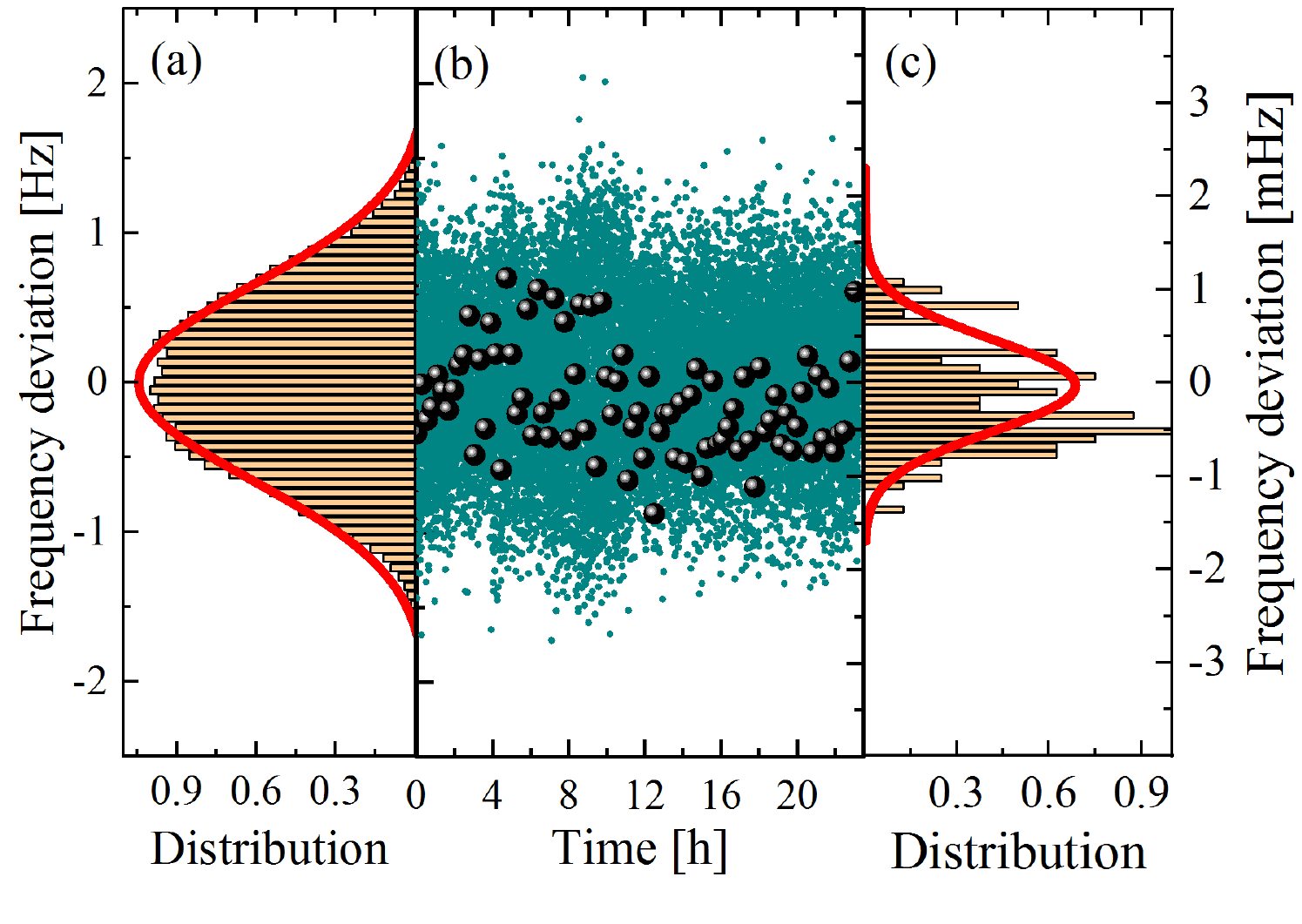}
\caption{Frequency comparison between input and access place frequencies after 70 km over the 100 km  fiber. (b) 84,014 data points were taken with dead-time free $\Pi$-type frequency counters with a 1 s gate time (green points, left axis). We calculated unweighted mean ($\Pi$-type) values for all cycle-slip free 1,000 s long segments, resulting in 84 data points (black dots, right frequency axis, enlarged scale). Histograms (brown bars) and Gaussian fits (red curves) for (a) frequency values taken with one second gate time and (c) 84 phase coherent 1,000-second frequency averages.}
\label{fig4}
\end{figure}

Complementary to the characterization of the stability and phase noise, the accuracy has to be throughout-fully examined by calculating the mean value of the end-to-end beatnote frequency offset. Figure \ref{fig4}(b) shows the frequency deviation of the beatnote’s data for the 70/30 km access place, recorded with a 1 s gate time and $\Pi$-type counters, over successive 84,014 s (green point, left axis) and the arithmetic mean of all cycle-slip free 1,000 s intervals (black dots, right axis). Histograms (brown bars) and Gaussian fits (red curves) of the frequency deviation for the access place after 70 km are also illustrated in Fig. \ref{fig4}(a) and (c). According to the Gaussian fit in Fig. \ref{fig4}(c), the calculated results demonstrate that the mean frequency is shifted by -28.9 $\mu$Hz ($-1.5\times10^{-19}$). The standard deviation of the 1,000 s data points is 493 $\mu$Hz ($2.5\times10^{-18}$), which is a factor of 1,000 smaller than the ADEV at 1 s as expected for this $\Pi$-type evaluation. Considering the long-term stability of frequency transfer as illustrated in Fig. \ref{fig3}, we conservatively estimate the accuracy of the transmitted optical signal as shown in the last data point of the ADEV, resulting in a relative frequency accuracy of $4.1\times10^{-19}$.

Adopting the same procedure, the mean frequency offset and the standard deviation for the 50/50 km place were calculated using the 1000 s point with the total 89,690 $\Pi$-type counter data to be 52.2 $\mu$Hz ($2.7\times10^{-19}$) and 454 $\mu$Hz ($2.3\times10^{-18}$), respectively. Taking into account the long-term ADEV at 10,000 s of the data set for the 50/50 km access place of $3.3\times10^{-19}$, we conservatively estimate that the mean frequency offset is $2.7\times10^{-19}$  with a statistical uncertainty of $3.3\times10^{-19}$  for the 50/50 km access place. We can conclude that there is no systematic frequency shift arising in the all-passive multiple-place phase noise cancellation setup at a level of a few $10^{-19}$.

In summary, we have presented a new method, making a stable optical frequency available at any arbitrary access places along the fiber link with passive phase noise cancellation. {In comparison with previous work, here we demonstrated that a high performance multiple-place optical frequency transfer over the widely adopted bus fiber topology with the open-loop design, mitigating some technical difficulties in conventional active multiple-access phase noise cancellation.} We experimentally demonstrate transferring of optical frequency to two different access places. After being compensated, delivering an optical frequency to different places with the relative frequency instability in terms of ADEV measured by the $\Pi$-mode frequency counter can be as low as $3.1\times10^{-15}$ at 1 s and $ 4.1\times10^{-19}$ at 10,000 s. The frequency uncertainty of the light after transferring through the fiber relative to that of the input light is a few  $10^{-19}$ for the access places over the 100 km fiber link. 

\medskip
\noindent\textbf{Funding.} This research was supported by the National Natural Science Foundation of China (NSFC) (61905143, 61627817).

\medskip
\noindent\textbf{Disclosures.} The authors declare no conflicts of interest.

%Manual citation list
%\begin{thebibliography}{1}
%\bibitem{Zhang:14}
%Y.~Zhang, S.~Qiao, L.~Sun, Q.~W. Shi, W.~Huang, %L.~Li, and Z.~Yang,
 % \enquote{Photoinduced active terahertz metamaterials with nanostructured
  %vanadium dioxide film deposited by sol-gel method,} Opt. Express \textbf{22},
  %11070--11078 (2014).
%\end{thebibliography}

% Please include bios and photos of all authors for aop articles
\ifthenelse{\equal{\journalref}{aop}}{%
\section*{Author Biographies}
\begingroup
\setlength\intextsep{0pt}
\begin{minipage}[t][6.3cm][t]{1.0\textwidth} % Adjust height [6.3cm] as required for separation of bio photos.
  \begin{wrapfigure}{L}{0.25\textwidth}
    \includegraphics[width=0.25\textwidth]{john_smith.eps}
  \end{wrapfigure}
  \noindent
  {\bfseries John Smith} received his BSc (Mathematics) in 2000 from The University of Maryland. His research interests include lasers and optics.
\end{minipage}
\begin{minipage}{1.0\textwidth}
  \begin{wrapfigure}{L}{0.25\textwidth}
    \includegraphics[width=0.25\textwidth]{alice_smith.eps}
  \end{wrapfigure}
  \noindent
  {\bfseries Alice Smith} also received her BSc (Mathematics) in 2000 from The University of Maryland. Her research interests also include lasers and optics.
\end{minipage}
\endgroup
}{}

\end{document}